\documentclass[10pt,a4paper]{article}

\RequirePackage[OT1]{fontenc}
\RequirePackage{amsthm,amsmath}
\RequirePackage[numbers]{natbib}
\RequirePackage[colorlinks,citecolor=blue,urlcolor=blue]{hyperref}

\theoremstyle{plain}
\newcommand\D{{\mathcal D}}
\newcommand\limn{\lim_{n\rightarrow\infty}}
\def\liminf{\mathop{\underline{\mathrm{lim}}}}
\newcommand\liminfn{\liminf_{n\rightarrow\infty}}
\newtheorem{thm}{Theorem}

\title{A note on the best attainable rates of convergence for estimates of the shape parameter of regular variation}

\author{Meitner Cadena\thanks{UPMC Paris 6 \& CREAR, ESSEC Business School;\, E-mail: meitner.cadena@etu.upmc.fr or b00454799@essec.edu or meitner.cadena@gmail.com}} 

\begin{document}

\maketitle

\begin{abstract}
Hall and Welsh gave in 1984 the lowest bound so far to rates of convergence for estimates of the shape parameter of regular variation.
We show that this bound 
can be improved.
\end{abstract}


\emph{Classification}: 62G05, 62G20

\vspace{2mm}

%
%


Hall and Welsh (hereafter HW) gave in 1984 a first lower bound of the accuracy of tail index estimation for a large class of distributions.
Since then, this result has been a reference to evaluate rates of convergence for other estimators of this parameter, and 
has motivated 
extensions of it 
for other classes of distributions (say e.g. \cite{BeirlantBouquiauxWerker2006}, \cite{Drees1998}, \cite{Novak2014}, 
\cite{Pfanzagl2000}, and \cite{Smith1987}).

Let $F$ be a differentiable distribution function (df) defined on the positive half-line such that, for positive constants $\alpha$, $\beta$, $C$ and $A$,
\begin{equation}\label{eq:20150805:01}
F'(x)=C\alpha x^{\alpha-1}(1+r(x))
\quad\textrm{where}\quad
\big|r(x)\big|\leq Ax^\beta,
\end{equation}
as $x\to0^+$.

Considering this type of dfs, HW \cite{HallWelsh1984} showed in 1984 that no estimator of $\alpha$ converges at a faster rate than $n^{-\beta/(2\beta+\alpha)}$ 
on certain neighborhoods of Pareto distributions (see e.g. \cite{Smith1987} or \cite{Drees1998}).
More precisely, these authors defined classes $\D=\D(\alpha_0,C_0,\epsilon,\rho,A)$ of dfs $F$ satisfying (\ref{eq:20150805:01}) and, in addition, $\big|\alpha-\alpha_0\big|\leq\epsilon$, 
$\big|C-C_0\big|\leq\epsilon$ and $\rho=\beta\big/\alpha$ for some given positive constants $\alpha_0$, $C_0$, $\epsilon$ and $A$. Let $\beta_0=\rho\alpha_0$.
Then it was shown that (see Theorem 1 in \cite{HallWelsh1984}), if $\alpha_n$ is an estimator of $\alpha$, constructed out of a random $n$-sample $X_1$, \ldots, $X_n$, satisfying
\begin{equation}\label{eq:20150805:02}
\liminfn\inf_{F\in\D}P\big(\big|\alpha_n-\alpha\big|\leq a_n\big)=1,
\end{equation}
then
%
$$
\liminfn n^{\beta_0/(2\beta_0+\alpha_0)}a_n=\infty.
$$
%
We find that the proof 
of this result, developed by HW, allows one, after some 
adequate modifications, to also prove

\begin{thm}\label{teo:20150805:01}
Suppose that for some $\alpha_0$, $C_0$, $\epsilon$ and $\rho$, we have (\ref{eq:20150805:02}) for all $A>0$. Then, for all 
$\nu\geq\beta_0\big/(2\beta_0+\alpha_0)$,
%
$$
\liminfn n^{\nu}a_n=\infty.
$$
%
\end{thm}

This means that $n^{-\beta/(2\beta+\alpha)}=n^{-\beta_0/(2\beta_0+\alpha_0)}=n^{-\rho/(2\rho+1)}$ because of $\rho=\beta\big/\alpha=\beta_0\big/\alpha_0$, is no longer a lower bound to convergence rates for estimators of shape parameters in distributions with regularly varying tails, as claimed by Theorem 1 given in \cite{HallWelsh1984}.

%

The proof of Theorem \ref{teo:20150805:01} is the same given by HW to prove Theorem 1 in \cite{HallWelsh1984}, but redefining conveniently two parameters.
We present these redefinitions and show how with these changes the original proof can still be applied.

In order to have a self-contained paper, we copy almost all of the proof of Theorem 1 in \cite{HallWelsh1984}.
The main changes in that proof are pointed out.

Let $\nu\geq\beta_0\big/(2\beta_0+\alpha_0)$.

For proving Theorem 1 in \cite{HallWelsh1984}, HW started constructing two densities $f_0$ and $f_1$, the first governed by fixed parameters $\alpha_0$, $C_0$ and the second by varying parameters $\alpha_1$, $C_1$, $C_2$, where $\alpha_1 = \alpha_0 + \tilde{\gamma}$, $\tilde{\gamma} = \lambda n^{-\nu}$, $\lambda > 0$, $\beta_1 = \rho\alpha_1$ and both $C_1$, $C_2$ $\to$ $C_0$ as $n\to\infty$.

Here we point out that we use $\tilde{\gamma}$ instead of $\gamma$. HW used $\gamma$ in the proof of Theorem 1 in \cite{HallWelsh1984}, where these authors defined it as $\gamma = \lambda n^{-\beta_1/(2\beta_1+\alpha_1)}$.

Specifically, HW defined

$$
f_0(x)=C_0\alpha_0x^{\alpha_0-1},\quad 0\leq x\leq C_0^{-1/\alpha_0},
$$

and

$$
f_1(x)=\left\{
\begin{array}{ll}
C_1\alpha_1x^{\alpha_1-1}+\Delta(x), & 0\leq x\leq\tilde{\delta} \\
C_2\alpha_0x^{\alpha_0-1}, & \tilde{\delta}<x\leq C_0^{-1/\alpha_0}.
\end{array}
\right.
$$

where $\tilde{\delta}=n^{-\nu/\beta_1}$, $k=\alpha_1+\beta_1-1$ and

$$
\Delta(x)=\left\{
\begin{array}{ll}
x^k, & 0< x\leq\tilde{\delta}\big/4 \\
\big(\tilde{\delta}\big/2-x\big)^k, & \tilde{\delta}\big/4< x\leq\tilde{\delta}\big/2 \\
-\big(x-\tilde{\delta}\big/2\big)^k, & \tilde{\delta}\big/2< x\leq3\tilde{\delta}\big/4 \\
-\big(\tilde{\delta}-x\big)^k, & 3\tilde{\delta}\big/4< x\leq\tilde{\delta}.
\end{array}
\right.
$$

Here we point out that we use $\tilde{\delta}$ instead of $\delta$. HW used $\delta$ in the proof of Theorem 1 in \cite{HallWelsh1984}, where these authors defined it as $\delta = n^{-1/(2\beta_1+\alpha_1)}$.


One can note that $\Delta(x)$ is continuous on $\big[0;\tilde{\delta}\big]$, that $\Delta(0) = \Delta(\tilde{\delta}) = 0$ and

$$
\int_0^{\tilde{\delta}}\Delta(x)\,dx=0.
$$

HW chose the constants $C_1$, $C_2$ so that for large $n$, $f_1$ is a proper, continuous density on $\big[0; C_0^{-1/\alpha_0}\big]$; that is,

\begin{equation}\label{(2.1)}
C_1\alpha_1\tilde{\delta}^{\alpha_1}=C_2\alpha_0\tilde{\delta}^{\alpha_0}
\end{equation}

and

\begin{equation}\label{(2.2)}
C_1\tilde{\delta}^{\alpha_1}+C_2\big(C_0^{-1}-\tilde{\delta}^{\alpha_0}\big)=1.
\end{equation}

Note that from (\ref{(2.1)})
$$
\limn\left(C_1-C_2\right)=
\limn C_2\left(\frac{\alpha_0}{\alpha_1}\tilde{\delta}^{-\tilde{\gamma}}-1\right)=
0
$$
and from (\ref{(2.1)}) and (\ref{(2.2)})
\begin{equation}\label{eq001}
C_2-C_0=C_0\left(C_2\tilde{\delta}^{\alpha_0}-C_1\tilde{\delta}^{\alpha_1}\right)=C_0\left(C_1\frac{\alpha_1}{\alpha_0}\tilde{\delta}^{\alpha_1}-C_1\tilde{\delta}^{\alpha_1}\right)=
\frac{C_0C_1}{\alpha_0}\tilde{\gamma}\tilde{\delta}^{\alpha_1},
\end{equation}
which gives
$$
\limn\left(C_2-C_0\right)=
\limn \frac{C_0C_1}{\alpha_0}\tilde{\delta}^{\alpha_1}\tilde{\gamma}=0.
$$
This guarantees that $C_1,C_2\to C_0$ as $n\to\infty$, as required for $C_1$ and $C_2$.

Then, the proof given by HW consisted initially of showing that
%

%
\begin{equation}\label{(2.3)}
\int_0^{C_0^{-1/\alpha_0}}\big(f_0(x)-f_1(x)\big)^{-2}f_0(x)dx=O\big(n^{-1}\big)
\end{equation}
as $n\to\infty$, and for all large $n$,
\begin{equation}\label{(2.4)}
f_1\in\D(\alpha_0,C_0,\epsilon,\rho,A).
\end{equation}
Note that trivially $f_0\in\D$.
The symbol $K$ denotes a positive generic constant.

By (\ref{eq001}), as $n\to\infty$,
\begin{equation}\label{(2.7)}
\big|C_2-C_0\big|=O\big(\tilde{\gamma}\tilde{\delta}^{\alpha_1}\big).
\end{equation}

We also have, as $n\to\infty$,
\begin{eqnarray}
\lefteqn{(2\alpha_1-\alpha_0)\int_0^{\tilde{\delta}}\left(C_0\alpha_0x^{\alpha_0-1}-C_1\alpha_1x^{\alpha_1-1}\right)^2\big(x^{\alpha_0-1}\big)^{-1}dx} \nonumber \\
 & = & (2\tilde{\gamma}+\alpha_0)\int_0^{\tilde{\delta}}\left(C_0^2\alpha_0^2x^{\alpha_0-1}-2C_0C_1\alpha_0\alpha_1x^{\alpha_1-1}+C_1^2\alpha_1^2x^{2\alpha_1-\alpha_0-1}\right)dx \nonumber \\
 & = & O\big(\tilde{\delta}^{\alpha_1}\big(C_0-C_1\tilde{\delta}^{\tilde{\gamma}}\big)^2+\tilde{\gamma}^2\tilde{\delta}^{\alpha_1}\big); \label{(2.8)}
\end{eqnarray}

using (\ref{eq001})

\begin{eqnarray}
\lefteqn{C_0-C_1\tilde{\delta}^{\tilde{\gamma}}\quad=\quad C_0-C_2+\tilde{\delta}^{-\alpha_0}\big(C_2C_0^{-1}-1\big)
\quad=\quad\big(C_0-C_2\big)\big(1-C_0^{-1}\tilde{\delta}^{-\alpha_0}\big)} \nonumber \\
 & = & -\frac{C_0C_1}{\alpha_0}\tilde{\gamma}\tilde{\delta}^{\alpha_1}\big(1-C_0^{-1}\tilde{\delta}^{-\alpha_0}\big) 
 \quad=\quad \frac{C_1}{\alpha_0}\tilde{\gamma}\big(\tilde{\delta}^{\tilde{\gamma}}-C_0\tilde{\delta}^{\alpha_1}\big) 
 \quad=\quad O\big(\tilde{\gamma}\big); \label{(2.8bis)}
\end{eqnarray}

and

\begin{equation}\label{(2.8bisbis)}
\int_0^{\tilde{\delta}}\big(\Delta(x)\big)^2x^{-\alpha_0+1}dx\quad\leq\quad K\int_0^{\delta}x^{2k-\alpha_0+1}dx\quad=\quad
O\big(\tilde{\delta}^{\alpha_1+2\beta_1}\big).
\end{equation}

Next, observing that

\begin{eqnarray*}
\lefteqn{\int_0^{C^{-1/\alpha_0}}\big(f_0(x)-f_1(x)\big)^2\big(f_0(x)\big)^{-1}dx} \\
 & = & \int_0^{\tilde{\delta}}\big(C_0\alpha_0x^{\alpha_0-1}-C_1\alpha_1x^{\alpha_1-1}-\Delta(x)\big)^2\big(C_0\alpha_0x^{\alpha_0-1}\big)^{-1}dx \\
 & \quad & +\int_{\tilde{\delta}}^{C_0^{-1/\alpha_0}}\big(C_0\alpha_0x^{\alpha_0-1}-C_2\alpha_0x^{\alpha_0-1}\big)^2\big(C_0\alpha_0x^{\alpha_0-1}\big)^{-1}dx \\
 & \leq & 2\int_0^{\tilde{\delta}}\left(C_0\alpha_0x^{\alpha_0-1}-C_1\alpha_1x^{\alpha_1-1}\right)^2\big(C_0\alpha_0x^{\alpha_0-1}\big)^{-1}dx \\
 & & + 2\int_0^{\tilde{\delta}}\big(\Delta(x)\big)^2\big(C_0\alpha_0x^{\alpha_0-1}\big)^{-1}dx 
+\big(C_0-C_2\big)^2\int_{\tilde{\delta}}^{C_0^{-1/\alpha_0}}C_0^{-1}\alpha_0x^{\alpha_0-1}dx,
\end{eqnarray*}

then, introducing (\ref{(2.7)}), combining (\ref{(2.8)}) and (\ref{(2.8bis)}), and using (\ref{(2.8bisbis)}), give

$$
\int_0^{C^{-1/\alpha_0}}\big(f_0(x)-f_1(x)\big)^2\big(f_0(x)\big)^{-1}dx\quad\leq\quad O\big(\tilde{\gamma}^{2}\tilde{\delta}^{\alpha_1}+\tilde{\delta}^{2\beta_1+\alpha_1}\big).
$$

(\ref{(2.3)}) immediately follows taking $\gamma$ and $\delta$ instead of $\tilde{\gamma}$ and $\tilde{\delta}$, as in the proof of Theorem 1 given in \cite{HallWelsh1984}.
Considering $\tilde{\gamma}$ and $\tilde{\delta}$,
we now have
$$
O\big(\tilde{\gamma}^{2}\tilde{\delta}^{\alpha_1}+\tilde{\delta}^{2\beta_1+\alpha_1}\big)
\quad=\quad O\big(\lambda^{2}n^{-2\nu-\nu\alpha_1/\beta_1}+n^{-\nu(2\beta_1+\alpha_1)/\beta_1}\big)
\quad=\quad O\big(n^{-\nu(2\beta_1+\alpha_1)/\beta_1}\big),
$$
and (\ref{(2.3)}) then follows too since
$$
\frac{\beta_1}{2\beta_1+\alpha_1}
\ \ =\ \ \frac{\beta_0}{2\beta_0+\alpha_0}
\ \ \leq\ \ \nu.
$$

The result (\ref{(2.4)}) will follows if we prove that
\begin{equation}\label{(2.9)}
\big|C_2\alpha_0x^{\alpha_0-1}-C_1\alpha_1x^{\alpha_1-1}\big|\ \ \leq\ \  Kx^{\alpha_1+\beta_1-1}
\end{equation}
uniformly in $\tilde{\delta}<x\leq C^{-1/\alpha_0}_0$ and large $n$.
By (\ref{eq001}),

\begin{eqnarray*}
\lefteqn{\big|C_2\alpha_0x^{\alpha_0-1}-C_0\alpha_0x^{\alpha_1-1}\big|
\quad=\quad\alpha_0x^{\alpha_0-1}\big|C_2-C_0\big|} \\
 & & =\quad\frac{C_0C_1}{\alpha_0}\tilde{\gamma}\tilde{\delta}^{\alpha_1}\alpha_0x^{\alpha_0-1}
\quad=\quad Kn^{-\nu}n^{-\nu(\alpha_1-\beta_1-\tilde{\gamma})/\beta_1}\tilde{\delta}^{\tilde{\gamma}+\beta_1}x^{\alpha_0-1}
\quad\leq \quad Kn^{-\nu\alpha_0/\beta_1}x^{\alpha_1+\beta_1-1}
\end{eqnarray*}

and so (\ref{(2.9)}) will follow if we show that for $\tilde{\delta}<x\leq C^{-1/\alpha_0}_0$,
\begin{equation}\label{(2.10)}
\big|C_0\alpha_0x^{\alpha_0-1}-C_1\alpha_1x^{\alpha_1-1}\big|\ \ \leq\ \  Kx^{\alpha_1+\beta_1-1}.
\end{equation}
But, using (\ref{(2.1)}) and (\ref{(2.2)}) gives, by $\tilde{\gamma}=\lambda\tilde{\delta}^{\beta_1}$,
$
C_0\alpha_0=C_0C_1\tilde{\delta}^{\alpha_1}\big(\alpha_0-\alpha_1\big)+C_1\alpha_1\tilde{\delta}^{\tilde{\gamma}},
$

\begin{eqnarray}
\big|C_0\alpha_0x^{\alpha_0-1}-C_1\alpha_1x^{\alpha_1-1}\big| & = & x^{\alpha_1-1}\big|C_0C_1\tilde{\delta}^{\alpha_1}\big(\alpha_0-\alpha_1\big)x^{-\tilde{\gamma}}+C_1\alpha_1\tilde{\delta}^{\tilde{\gamma}}x^{-\tilde{\gamma}}-C_1\alpha_1\big| \nonumber \\
 & \leq & K_1x^{\alpha_1-1}\tilde{\gamma}\tilde{\delta}^{\alpha_1-\tilde{\gamma}}+K_2x^{\alpha_1-1}\left|1-\big(\tilde{\delta}\big/x\big)^{\tilde{\gamma}}\right| \nonumber \\
 & \leq & K_3x^{\alpha_1+\beta_1-1}\tilde{\delta}^{\alpha_1}+K_4x^{\alpha_1-1}\tilde{\gamma}\,\log\big(x\big/\tilde{\delta}\big). \label{(2.11)}
\end{eqnarray}

Now, $x^{-\beta_1}\tilde{\gamma}\,\log\big(x\big/\tilde{\delta}\big)=\big(x\big/\tilde{\delta}\big)^{\beta_1}\,\log\big(x\big/\tilde{\delta}\big)$, and is maximized by taking
$x\big/\tilde{\delta}=e^{1/\beta_1}$.
Therefore by (\ref{(2.11)}),
$$
\big|C_0\alpha_0x^{\alpha_0-1}-C_1\alpha_1x^{\alpha_1-1}\big|\quad \leq\quad 
K_3x^{\alpha_1+\beta_1-1}\tilde{\delta}^{\alpha_1}+K_5x^{\alpha_1+\beta_1-1}\quad \leq\quad 
K_6x^{\alpha_1+\beta_1-1}
$$
uniformly in $\tilde{\delta}<x\leq C^{-1/\alpha_0}_0$.
This proves (\ref{(2.10)}), and completes the proof of (\ref{(2.4)}).

For what follows, the proof of Theorem 1 in \cite{HallWelsh1984} was inspired by Farrell (1982) \cite{Farrell1982}.
Observing that, using the Cauchy-Schwarz inequality,
\begin{eqnarray}
\lefteqn{P_{f_1}\left(\big|\alpha_n(X_1,\ldots,X_n)-\alpha_1\big|\leq a_n\right)} \nonumber \\
 & = & E_{f_0}\left[I\left(\big|\alpha_n(X_1,\ldots,X_n)-\alpha_1\big|\leq a_n\right)\prod_{i=1}^n\big(f_1(X_i)\big/f_0(X_i)\big)\right] \nonumber \\
 & \leq & \left(P_{f_0}\left(\big|\alpha_n(X_1,\ldots,X_n)-\alpha_1\big|\leq a_n\right)\right)^{1/2} \nonumber \\
 & & \times
\left(E_{f_0}\left[\prod_{i=1}^n\big(f_1(X_i)\big/f_0(X_i)\big)^2\right]\right)^{1/2}, \label{(2.12)}
\end{eqnarray}
and
\begin{eqnarray*}
\left(E_{f_0}\left[\prod_{i=1}^n\left(\frac{f_1(X_i)}{f_0(X_i)}\right)^2\right]\right)^{1/n}
 & = & \int_0^{C_0^{-1/\alpha_0}}\frac{\big(f_1(x)\big)^2}{f_0(x)}dx \\
 & = & 1+\int_0^{C_0^{-1/\alpha_0}}\left(f_1(x)-f_0(x)\right)^2\left(f_0(x)\right)^{-1}dx \\
 & = & 1+O(n^{-1}),
\end{eqnarray*}
using (\ref{(2.3)}).
Hence
\begin{equation}\label{(2.13)}
P_{f_1}\left(\big|\alpha_n(X_1,\ldots,X_n)-\alpha_1\big|\leq a_n\right)\ \ \leq\ \  K\,\left(P_{f_0}\left(\big|\alpha_n(X_1,\ldots,X_n)-\alpha_1\big|\leq a_n\right)\right)^{1/2}.
\end{equation}
By hypothesis and by (\ref{(2.4)}), the left-hand side of (\ref{(2.13)}) tends to 1 as $n\to\infty$.
Therefore $P_{f_0}\left(\big|\alpha_n(X_1,\ldots,X_n)-\alpha_1\big|\leq a_n\right)$ is bounded away from zero as $n\to\infty$.
Also by hypothesis, $P_{f_0}\left(\big|\alpha_n(X_1,\ldots,X_n)-\alpha_1\big|\leq a_n\right)$ tends to 1 as $n\to\infty$, and so
$$
P_{f_0}\left(\left\{\big|\alpha_n(X_1,\ldots,X_n)-\alpha_1\big|\leq a_n\right\}\cap\left\{\big|\alpha_n(X_1,\ldots,X_n)-\alpha_0\big|\leq a_n\right\}\right)
$$
is bounded away from zero. Consequently, for large $n$,
$$
\big|\alpha_1-\alpha_0\big|
\ \ \leq\ \ 2a_n
$$
%
%
%
that is, $\tilde{\gamma}=\lambda n^{-\nu}\leq2a_n$, and so
$$
\liminf_{n\to\infty} n^{\nu} a_n\ \ \geq\ \ \frac{\lambda}{2}.
$$
Since this is true for each $\lambda > 0$, Theorem \ref{teo:20150805:01} is proved.

\section*{Acknowledgments} 
The author gratefully acknowledges the support of SWISS LIFE through its ESSEC research program on 'Consequences of the population ageing on the insurances loss'.

\end{document}